\documentclass[12pt]{iopart}
\usepackage{comment}
\usepackage{iopams}
\usepackage{graphicx} 
\usepackage{xcolor}

\usepackage[
colorlinks=true,
    linkcolor={red!50!black},
    citecolor={blue!50!black},
    urlcolor={blue!80!black}]{hyperref}

\DeclareMathAlphabet\mathbfcal{OMS}{cmsy}{b}{n}

\begin{document}



\title[Flow equations for cold Bose gases]{Flow equations for cold Bose gases}

\author{A G Volosniev$^1$ and H-W Hammer$^1$$^,$$^2$}
\address{$^1$ Institut f{\"u}r Kernphysik, Technische Universit{\"a}t Darmstadt, 64289 Darmstadt, Germany}
\address{$^2$ ExtreMe Matter Institute EMMI, GSI Helmholtzzentrum f{\"u}r Schwerionenforschung GmbH, 64291 Darmstadt, Germany}
\ead{volosniev@theorie.ikp.physik.tu-darmstadt.de}

\begin{abstract}
  We derive flow equations for cold atomic gases with one
  macroscopically populated energy level. The generator is chosen such
  that the ground state decouples from all other states in the system as the
  renormalization group flow progresses. We propose a self-consistent
  truncation scheme for the flow equations at the level of three-body
  operators and show how
  they can be used to calculate the ground state energy of a general
  $N$-body system. Moreover, we provide a general method to estimate the
  truncation error in the calculated energies.
  Finally, we test our scheme by benchmarking to the
  exactly solvable Lieb-Liniger model and find good agreement for weak
  and moderate interaction strengths.
  
  \end{abstract}



\maketitle

\section{Introduction}

The worlds of many- and few-body physics are generally far apart. In the former, the number of particles is often infinite, while few-body
systems normally do not contain more than a handful of particles. The typical goal in many-body physics is to calculate thermodynamic
quantities such as the energy per particle and the density profile.  However, the large number of
degrees of freedom in many-body systems usually means that various approximations and/or large computational resources are needed to achieve this goal.
In contrast, it is often possible to solve few-body problems exactly, i.e., to find the full spectrum of the Hamiltonian and the corresponding
wave functions. 
There is an interesting class of systems that are in between these two extremes. These are finite systems in which the
number of particles is sufficiently large for many-body phenomena, such as superfluidity or Bose-Einstein condensation, to
emerge~\cite{Ceperley1989,Grebenev1998,ketterle1996}; but they are still small enough to be  within reach for  numerically exact {\it ab-initio} calculations that use microscopic Hamiltonians.  The investigation of such finite systems is crucial to understand how many-body phenomena
arise from few-body body physics and microscopic interactions of the constituents.

To investigate this progression from few- to many-body behavior theoretically one needs reliable numerical techniques in the transition
region. There exists a number of suitable techniques in physics and chemistry and new methods are being developed~(see, e.g., references~\cite{Bartlett2007,Lee:2008fa,Navratil:2009ut,Schollwock2011,Hagen:2013nca,Furnstahl:2013oba,carlson2015}).
A significant breakthrough was made with the development of flow equation methods which are also referred to as
the similarity renormalization group (SRG)~\cite{glazek1993,wegner1994}. In this approach, a set
of differential  equations is  solved to obtain unitarily equivalent Hamiltonians with desirable properties.
This set is determined by a generator, which controls the change of the Hamiltonian at every step of the evolution.  Note that this generator is determined dynamically. Its matrix elements depend on
the flow parameter $s$ and are calculated at every step of the evolution from the transformed Hamiltonian. This represents one of the key advantages of the SRG, which allows one to find a  (block-)diagonal representation of the Hamiltonian. 

Recently a new approach based on flow equations, the in-medium similarity renormalization group (IMSRG),
has been proposed for nuclear physics problems where the fundamental degrees of freedom are fermionic~\cite{Tsukiyama2011}.
The IMSRG is a very promising  method for
medium-mass nuclei, which lie exactly in the few- to many-body transition region discussed above
(see~\cite{hergert2016} for a recent review). 

In this paper, we develop a similar method for cold Bose gases. To this end, we write flow equations for bosonic systems with a macroscopic occupation of one state. We introduce a suitable truncation scheme that facilitates numerical calculations and discuss its accuracy.
In particular, we provide an algorithm to estimate the truncation error using perturbation theory.
We validate our method using the exactly solvable Lieb-Liniger model in one dimension, and show that even without preliminary knowledge
of the reference state our method can be used to accurately describe systems with weak and intermediate interaction strengths.  
   
The paper is organized as follows: in section~\ref{sec:prel}, we review the foundations of the SRG method. In section~\ref{sec:flow_eq}, we
introduce the Hamiltonian of interest and write down the flow equations to find its eigenvalue. Here we also discuss the accuracy of our approach and provide a way to estimate the accuracy of the calculated energies.
We test the method in section~\ref{sec:ll_model} using the exactly solvable Lieb-Liniger model as a benchmark. Section~\ref{sec:concl} concludes the paper with a summary of our results and a brief outlook on the generalization to three spatial dimensions.   For the reader's convenience, we include six appendices
with technical details on the evaluation of commutators, the truncation of the three-body operator, the
convergence of the two-body energy, the effective interaction used in the Lieb-Liniger model, the use of White-type generators, and the error estimation.

\section{Preliminaries}
\label{sec:prel}

For a self-contained discussion, we first
review the SRG method as it forms the basis of our approach
(cf.~\cite{glazek1993, wegner1994,hergert2016,  white2002, kehrein2006}). To this end,
we introduce a real symmetric matrix ${\bf M}$ that represents a linear operator in a particular
basis\footnote{Two comments are in order here. First, we use bold
  type for matrices and operators, e.g., ${\bf M}$, and italic type for the corresponding matrix
  elements, e.g., $M_{ij}$. Second, we choose to work with a real matrix  ${\bf M}$
  to simplify the discussion.
  The ideas presented here can be extended straightforwardly to Hermitian matrices.}.
If we transform this basis using some orthogonal matrix ${\bf Q}$ (i.e., ${\bf QQ}^T={\bf I}$,
where ${\bf I}$ is the identity matrix) then the linear operator will be represented by the new
matrix ${\bf M}({\bf Q})\equiv{\bf Q M Q}^T$, which is unitarily equivalent to $\mathbf{M}$. The
SRG equations simply describe the change of ${\bf M}$ for a small change of the basis:
${\bf Q}={\bf I}+ {\boldsymbol \eta}\delta s$ ($|\delta s|\ll 1$),
\begin{equation}
  {\bf M}({\bf Q}) = {\bf M} + [{\boldsymbol \eta},{\bf M}]\delta s +\ldots\,.
\end{equation}
In the limit $\delta s\to 0$, the SRG equations can be written in the differential form: 
\begin{equation}
\frac{\mathrm{d}M_{ij}}{\mathrm{d}s} = \sum_{k}(\eta_{ik}M_{kj}-M_{ik}\eta_{kj}).
\label{eq:flow_system}
\end{equation}
They  define the evolution of matrix elements $M_{ij}$ driven by
the skew-symmetric matrix ${\boldsymbol \eta}=-{\boldsymbol \eta}^T$.
By specifying ${\boldsymbol \eta}$, one finds a unitarily equivalent to ${\bf M}$ matrix  with some desired properties.
Note that the system of equations~(\ref{eq:flow_system}) is often called the ``flow equation'',
as it defines the ``flow'' of matrix elements under the SRG transformation, and the ``generator'' ${\boldsymbol \eta}$
determines the flow by defining the ``direction'' of the transformation at each value of $s$.   


We illustrate the evolution using a generator ${\boldsymbol \eta}$ that contains only two non-zero elements $\eta_{ab}=-\eta_{ba}$,
i.e., $\eta_{ik}(s)=\alpha(s)(\delta_{ia}\delta_{kb}-\delta_{ib}\delta_{ka})$. This matrix leads to the system of equations
\begin{equation}
\frac{\mathrm{d}M_{ij}}{\mathrm{d}s} = \alpha(M_{ib}\delta_{ja}+M_{jb}\delta_{ia}-M_{ia}\delta_{bj}-M_{ja}\delta_{bi}),
\end{equation}
in which the element $M_{ab}=M_{ba}$ is transformed as
\begin{equation}
\frac{\mathrm{d}M_{ab}}{\mathrm{d}s}=\frac{\mathrm{d}M_{ba}}{\mathrm{d}s}=\alpha(M_{bb}-M_{aa}).
\label{eq:Mab}
\end{equation}
Let us assume that we want the flow to eliminate the element $M_{ab}$ as $s\to\infty$, e.g., by demanding
that $M_{ab}(s)=M_{ab}(0)e^{-s}$. Inserting this ansatz into~(\ref{eq:Mab}) we find that $\alpha=-M_{ab}/(M_{bb}-M_{aa})$
fulfills this requirement\footnote{Note that to eliminate $M_{ab}$, we could also have chosen $\alpha(s)=-M_{ab}f(s)/(M_{bb}-M_{aa})$
with $f(s)>0$, e.g., $f(s)=|M_{bb}-M_{aa}|$, as then $\frac{\mathrm{d} M^2_{ab}}{\mathrm{d}s}< 0$ if $M_{ab}\neq 0$, which means that $M_{ab}(s)$ dies off.}, i.e., it decouples the basis states with numbers $a$ and $b$. Note, however, that to achieve this decoupling,
the flow usually needs to couple states that were not coupled before. For example, if we had $M_{cd}=0$ at $s=0$, then this element
will attain a non-zero value if $M_{cb}\delta_{da}+M_{db}\delta_{ca}-M_{ca}\delta_{bd}-M_{da}\delta_{bc}\neq 0$.

Let us give another example of how one can obtain a new matrix with some desired properties by choosing an appropriate generator
${\boldsymbol \eta}$.  
To this end, we use a generator that contains only one row and one column, i.e.,
$\eta_{ik}=\delta_{i0}\alpha_k-\delta_{k0}\alpha_i$ with $\alpha_0=0$. The corresponding flow equations are
\begin{eqnarray}
\frac{\mathrm{d}M_{0\,i>0}}{\mathrm{d}s}=-M_{00}\alpha_i+\sum_{k}\alpha_k M_{ki}, \\
\frac{\mathrm{d}M_{i>0\,j>0}}{\mathrm{d}s}=-M_{0j}\alpha_i-M_{i0}\alpha_j.
\end{eqnarray}
The prescription $\alpha_{i>0}=-M_{0i}$, which is inspired by the previous example, leads to 
\begin{equation}
\frac{\mathrm{d}M_{0i>0}}{\mathrm{d}s}=-\sum_{k\neq0} (M_{ik}-M_{00}I_{ik})M_{0k}.
\end{equation}
A formal solution to this equation can be found using the Magnus expansion
\begin{equation}
M_{0i}(s)=\sum_{k\neq 0}\left(\mathbfcal {T} e^{-\int_0^s ({\bf M}-M_{00}{\bf I})'\mathrm{d}s}\right)_{ik}M_{0k}(0),
\end{equation}
here $\mathbfcal{T}$ denotes the $s$-ordering operator (see, e.g.,~\cite{kehrein2006, blanesa2009}), and $'$ means that the first row and the first
column should be crossed out from the matrix. If all $M_{0j}$ are initially small (i.e., much smaller than the
differences of the eigenvalues of $\bf{M}$), then  the long time behavior can be estimated by examining the
matrix $ ({\bf M}(0)-M_{00}(0){\bf I})'$. This shows that if $M_{00}(0)$ is close to the ground state then
$M_{0i}$ is driven to zero during the evolution, and hence $M_{00}(s\to\infty)$ is the ground state of the matrix.
These considerations can be useful in physics problems, as they allow one to find  eigenenergies of a
system by diagonalizing (block-diagonalizing) the corresponding Hamiltonian. This statement will be exemplified below.

\section{Flow Equations}
\label{sec:flow_eq}

\subsection{Hamiltonian}

We now consider a system of $N$ bosons that is described by the Hamiltonian
\begin{equation}
  {\bf H}= A_{ij}{\boldsymbol a}_i^\dagger {\boldsymbol a}_j + \frac{1}{2}B_{ijkl}{\boldsymbol a}_i^\dagger
  {\boldsymbol a}_j^\dagger {\boldsymbol a}_k {\boldsymbol a}_l,
\label{eq:ham}
\end{equation}
where ${\boldsymbol a_{\alpha_1}}$ is the standard annihilation operator\footnote{From now on  we adopt in the numbered equations
  the Einstein summation convention for the letters from the Latin alphabet, i.e.,
  $A_{ij}{\boldsymbol a}_i^\dagger {\boldsymbol a}_j \equiv \sum_{ij} A_{ij}{\boldsymbol a}_i^\dagger {\boldsymbol a}_j$,
  and reserve the indices $\alpha_{1,2,...}$ for the places where this convention is not implied.}.
Since the system is bosonic, ${\bf H}$ is symmetrized with respect to particle exchanges, i.e., $B_{ijkl}=B_{jikl}=B_{ijlk}$.
For our numerical calculations this Hamiltonian should be written as a finite-dimensional matrix. Therefore,
we assume that the sums in every index run only up to some number $n$ that defines the dimension of the used one body basis.  

We are mainly interested in the ground state properties of systems with a macroscopic population of one state (condensate).
To incorporate our intentions in the Hamiltonian, we normal order operators using the reference state
$\Phi=\prod_{\alpha=1}^N f(x_\alpha)$, where $f(x)$ is some one body function that approximates the condensate
(e.g., obtained by solving a suitable Gross-Pitaevski equation):
\begin{eqnarray}
:{\boldsymbol a}_{\alpha_1}^\dagger {\boldsymbol a}_{\alpha_2}: & = {\boldsymbol a}_{\alpha_1}^\dagger {\boldsymbol a}_{\alpha_2} - {\bf I}\rho_{\alpha_1\alpha_2},  \\
:{\boldsymbol a}_{\alpha_1}^\dagger {\boldsymbol a}_{\alpha_2}^\dagger {\boldsymbol a}_{\alpha_3} {\boldsymbol a}_{\alpha_4}: &= 
{\boldsymbol a}_{\alpha_1}^\dagger {\boldsymbol a}_{\alpha_2}^\dagger {\boldsymbol a}_{\alpha_3} {\boldsymbol a}_{\alpha_4}-{\bf I} \rho_{\alpha_1\alpha_2\alpha_3\alpha_4} \nonumber \\ &- \kappa(1+{\bf P}_{\alpha_1\alpha_2})(1+{\bf P}_{\alpha_3\alpha_4})\rho_{\alpha_2\alpha_3} :{\boldsymbol a}_{\alpha_1}^\dagger {\boldsymbol a}_{\alpha_4}:,
\end{eqnarray}
where ${\bf P}_{\alpha_1\alpha_2}$ exchanges the indices $\alpha_1$ and $\alpha_2$, $\kappa\equiv\frac{N-1}{2N}$,
$\rho_{\alpha_1\alpha_2}\equiv\langle \Phi|{\boldsymbol a}_{\alpha_1}^\dagger {\boldsymbol a}_{\alpha_2}|\Phi \rangle$,
$\rho_{\alpha_1\alpha_2\alpha_3\alpha_4}\equiv\langle \Phi|{\boldsymbol a}_{\alpha_1}^\dagger
{\boldsymbol a}_{\alpha_2}^\dagger {\boldsymbol a}_{\alpha_3} {\boldsymbol a}_{\alpha_4}|\Phi \rangle$.
These operators connect the reference state to the states that contain one and two excitations respectively.
 
Using the normal-ordered operators we rewrite the Hamiltonian as
\begin{eqnarray}
  {\bf H}=\epsilon N {\bf I} +f_{ij} :{\boldsymbol a}_i^\dagger {\boldsymbol a}_j: + \frac{1}{2}\Gamma_{ijkl}:{\boldsymbol a}_i^\dagger
  {\boldsymbol a}_j^\dagger {\boldsymbol a}_k {\boldsymbol a}_l:  , 
\end{eqnarray}
where 
\begin{eqnarray}
\epsilon N&=A_{ij}\rho_{ij} + \frac{1}{2}B_{ijkl}\rho_{ijkl},  \\
f_{\alpha_1 \alpha_2} &= A_{\alpha_1 \alpha_2} +  2\kappa B_{\alpha_1 i j\alpha_2}\rho_{ij}, \\
\Gamma_{\alpha_1 \alpha_2\alpha_3\alpha_4}&=B_{\alpha_1 \alpha_2\alpha_3\alpha_4},
\end{eqnarray}
$\epsilon$ is the energy per particle in the reference state, and the elements $f_{\alpha_1 \alpha_2}$ and
$\Gamma_{\alpha_1 \alpha_2\alpha_3\alpha_4}$ describe one- and two-body excitations, correspondingly. We will construct
the Hamiltonian matrix using the basis that contains $f(x)$ as the zero element, therefore, from now on we use
$\rho_{\alpha_1 \alpha_2}=\delta_{\alpha_10}\delta_{\alpha_20}N$ and $\rho_{\alpha_1\alpha_2\alpha_3 \alpha_4}=
\delta_{\alpha_1 0}\delta_{\alpha_2 0}\delta_{\alpha_3 0}\delta_{\alpha_4 0}N(N-1)$.

\subsection{Truncated flow equations}

Our goal is to find a matrix representation of ${\bf H}$ in which the couplings to the reference state vanish, i.e., $f_{i0}=\Gamma_{ij00}=0$,
so $\epsilon$ is an eigenenergy. To achieve this, we write ${\bf H}$ in a particular basis and then use the flow equations
\begin{equation}
\frac{\mathrm{d}{\bf H}(s)}{\mathrm{d}s} = [{\boldsymbol \eta}(s),{\bf H}(s)],
\label{eq:flow_eq_ham}
\end{equation}
where the antihermitian matrix ${\boldsymbol\eta}$ eliminates the couplings.
To solve this equation, we assume that during the flow the generator and the Hamiltonian contain only one- and two-body operators, i.e.,
\begin{eqnarray}
\label{eq:ham_eta}
{\boldsymbol \eta}(s)&=\xi_{ij}(s) :{\boldsymbol a}_i^\dagger {\boldsymbol a}_j: + \frac{1}{2}\eta_{ijkl}(s) :{\boldsymbol a}_i^\dagger {\boldsymbol a}_j^\dagger {\boldsymbol a}_k {\boldsymbol a}_l:,  \\
{\bf H}(s)&=\epsilon(s) N {\bf I}+f_{ij}(s) :{\boldsymbol a}_i^\dagger {\boldsymbol a}_j: + \frac{1}{2}\Gamma_{ijkl}(s):{\boldsymbol a}_i^\dagger {\boldsymbol a}_j^\dagger {\boldsymbol a}_k {\boldsymbol a}_l:. 
\label{eq:eta_ham}
\end{eqnarray}
 For now we leave the parameters $\xi_{ij}$ and $\eta_{ijkl}$ undetermined. We just mention that they must be chosen such that the couplings vanish at $s\to\infty$. This is usually achieved by calculating $\xi_{ij}(s)$ and $\eta_{ijkl}(s)$ for every $s$ from the evolved matrix elements of the Hamiltonian.
We give a possible choice of ${\boldsymbol \eta}$ in the next section. It is worthwhile noting that since ${\boldsymbol \eta}$ is antihermitian,
the following relations must be satisfied
$\xi_{ji}=-\xi_{ij}^{*}$, and $\eta_{klij}=-\eta_{ijkl}^{*}$. Moreover, we assume that $\eta_{ijkl}=\eta_{jikl}=\eta_{ijlk}$,
because by construction
\begin{eqnarray}
  :{\boldsymbol a}_{\alpha_1}^\dagger {\boldsymbol a}_{\alpha_2}^\dagger
{\boldsymbol a}_{\alpha_3} {\boldsymbol a}_{\alpha_4}: = :{\boldsymbol a}_{\alpha_2}^\dagger
{\boldsymbol a}_{\alpha_1}^\dagger {\boldsymbol a}_{\alpha_3} {\boldsymbol a}_{\alpha_4}: = :
{\boldsymbol a}_{\alpha_1}^\dagger {\boldsymbol a}_{\alpha_2}^\dagger {\boldsymbol a}_{\alpha_4} {\boldsymbol a}_{\alpha_3}:.  
\end{eqnarray}

Note that equations~(\ref{eq:flow_eq_ham}), (\ref{eq:ham_eta}) and (\ref{eq:eta_ham}) do not lead in a general case to a self-consistent system of equations.
Indeed, the commutator\footnote{From now on we omit the argument $s$ whenever it cannot cause confusion.} $[{\boldsymbol \eta},{\bf H}]$
contains the three body operator (see~\ref{app:comm})
\begin{equation}
[{\boldsymbol \eta},{\bf H}]^{(3)}=(\eta_{iklj}\Gamma_{jbcd}-\Gamma_{iklj}\eta_{jbcd}){\boldsymbol a}_i^\dagger {\boldsymbol a}_k^\dagger {\boldsymbol a}_b^\dagger {\boldsymbol a}_l {\boldsymbol a}_c {\boldsymbol a}_d,
\label{eq:three_body_term}
\end{equation} 
where the superscript $(3)$ corresponds to the piece of the commutator that contains three-body operators.
This piece is apparently beyond the scheme put forward in~(\ref{eq:ham_eta}) and should be omitted. To this end, we extract
from $[{\boldsymbol \eta},{\bf H}]^{(3)}$ the terms that contain at least one operator ${\boldsymbol a}_0^\dagger {\boldsymbol a}_0$,
and put to zero the remaining pieces (called ${\bf W}$). The operator ${\boldsymbol a}_0^\dagger {\boldsymbol a}_0$ is then treated as a
constant because of the assumed macroscopic occupation of the lowest state (see~\ref{app:trunc}).

After the three-body operator is truncated, we end up with a closed system of equations. To write it down, we equate the coefficients in front of the
same operators,~i.e.,  
\begin{eqnarray}
\fl \frac{\mathrm{d}\epsilon}{\mathrm{d}s} = S_{00} + (N-1)\left(\frac{1}{2}S_{00ii00}-S_{000000}\right)
\label{eq:flow_eq_bosons1},\\
\fl \frac{\mathrm{d}f_{\alpha_1 \alpha_2}(s)}{\mathrm{d}s} =- (N-1)^2(S_{0\alpha_100\alpha_20}+S_{0\alpha_1\alpha_2000}+S_{000\alpha_1\alpha_20}) \nonumber +(N-1)S_{\alpha_10ii0\alpha_2}\\ 
 +S_{0\alpha_1\alpha_20}+\frac{(N-1)(N-2)}{2}(S_{00\alpha_2\alpha_100}+S_{0\alpha_100\alpha_20}D_{\alpha_2}D_{\alpha_1})  \nonumber \\+ \frac{(N-1)(N-2)}{2}(S_{0\alpha_1\alpha_2000}D_{\alpha_1}+S_{000\alpha_1\alpha_20}D_{\alpha_2}) +  S_{\alpha_1\alpha_2},
\label{eq:flow_eq_bosons2} \\
\fl \frac{\mathrm{d}\Gamma_{\alpha_1\alpha_2\alpha_3\alpha_4}(s)}{\mathrm{d}s}=\frac{(1+P_{\alpha_1\alpha_2})(1+P_{\alpha_3\alpha_4})}{2}\bigg(S_{\alpha_1\alpha_2\alpha_3\alpha_4} \nonumber \\-(N-1)(S_{\alpha_1\alpha_2\alpha_300\alpha_4}+S_{\alpha_100\alpha_2\alpha_3\alpha_4}+\frac{1}{2}S_{\alpha_1\alpha_2ii\alpha_3\alpha_4})\nonumber \\ +(N-2)D_{\alpha_1}D_{\alpha_4}S_{0\alpha_1\alpha_3\alpha_2\alpha_40}+(N-2)I_{\alpha_1\alpha_2}D_{\alpha_4} S_{\alpha_1\alpha_2\alpha_300\alpha_4} \nonumber \\ + (N-2)D_{\alpha_1} I_{\alpha_3\alpha_4}S_{\alpha_100\alpha_2\alpha_3\alpha_4}+(N-2)I_{\alpha_1\alpha_2}I_{\alpha_3\alpha_4}S_{\alpha_1\alpha_2ii\alpha_3\alpha_4}\bigg),
\label{eq:flow_eq_bosons3}
\end{eqnarray}
where
$D_{\alpha_1}=2-\delta_{\alpha_1 0}$, $I_{\alpha_1 \alpha_2}=1+\delta_{\alpha_10}\delta_{\alpha_2 0}-2\delta_{\alpha_2 0}$, and
\begin{eqnarray}
S^{(1)}_{\alpha_1 \alpha_2} &= \xi_{\alpha_1 i}f_{i\alpha_2}-f_{\alpha_1 i}\xi_{i\alpha_2},  \nonumber \\
S^{(2)}_{\alpha_1 \alpha_2\alpha_3\alpha_4} &= \xi_{\alpha_1i}\Gamma_{i\alpha_2\alpha_3\alpha_4}-\Gamma_{\alpha_1\alpha_2\alpha_3i}\xi_{i\alpha_4}  + \eta_{\alpha_1\alpha_2\alpha_3i}f_{i\alpha_4}-f_{\alpha_1i}\eta_{i\alpha_2\alpha_3\alpha_4}, \nonumber \\
S_{\alpha_1\alpha_2\alpha_3\alpha_4\alpha_5\alpha_6} &= \eta_{\alpha_1\alpha_2\alpha_3i}\Gamma_{i\alpha_4\alpha_5\alpha_6}-\Gamma_{\alpha_1\alpha_2\alpha_3i}\eta_{i\alpha_4\alpha_5\alpha_6}. 
\end{eqnarray}
This system of equations can be solved using standard solvers of ordinary differential equations. During the evolution, an appropriate choice
of ${\boldsymbol \eta}$ eliminates the couplings $f_{i0}=0$ and $\Gamma_{ij00}=0$, so that $\epsilon(s\to\infty)$ approximates an eigenenergy
of the system. It is worthwhile noting that it is not guaranteed that $\epsilon$ is close to the ground state energy, unless  $\Phi$
describes the ground state wave function ``well'' (so that $f_{i0}$ and $\Gamma_{ij00}$ are much smaller than the differences of the eigenenergies of ${\bf H}$). 

\subsection{Error estimation}

Since the flow equations are truncated at the level of three-body operators and beyond, it is important to estimate the error induced by this
approximation. Let us imagine that we have integrated the flow equations (\ref{eq:flow_eq_bosons1})-(\ref{eq:flow_eq_bosons3}) up to $s\to\infty$, and obtained the operator ${\bf H}(s)$ within our truncation scheme as well as the generator ${\boldsymbol \eta}(s)$. Now we assume that ${\boldsymbol \eta}$ is fixed for every $s$ and use it to introduce the operator $\mathbfcal{H}$ that solves equation~(\ref{eq:flow_eq_ham}) with the initial condition $\mathbfcal{H} (s=0) = {\bf H}(s=0)$ without any truncations.
Hence, $\mathbfcal{H}$ is unitarily equivalent to ${\bf H}(0)$. We emphasize that ${\boldsymbol \eta}(s)$ is given from the beginning for every $s$ and not obtained dynamically as before.

The operator $\mathbfcal{H}$ can be written as $\mathbfcal{H}(s)={\bf H}(s)+{\bf H}_{a}(s)$, where ${\bf H}(s)$ is obtained from the truncated flow and ${\bf H}_{a}$ satisfies the equation
\begin{eqnarray}
\frac{\mathrm{d} {\bf H}_{a}(s)}{\mathrm{d}s}&={\bf W}(s)+[{\boldsymbol\eta(s)},{\bf H}_{a}(s)],
\label{eq:correction_flow}
\end{eqnarray} 
supplemented by the initial condition ${\bf H}_{a}(s=0)=0$. Note that the operator ${\bf H}_{a}$ is generated by ${\bf W}$, which is the part of~(\ref{eq:three_body_term}) that is
neglected in our truncation scheme. Therefore, we postulate that our approximation is meaningful only if ${\bf H_a}(s\to\infty)$ can be treated as a small perturbation for the state of interest. In this case $\epsilon(s\to\infty)$ is close to the exact eigenenergy of the operator $\mathbfcal{H}$. 

To estimate ${\bf H_a}(s)$ we write two formal solutions to~(\ref{eq:correction_flow})   
\begin{eqnarray}
{\bf H}_{a}(s)=\int_0^s{\bf W}\mathrm{d}x + \int_{0}^s[{\boldsymbol \eta}(x),{\bf H}_{a}(x)]\mathrm{d}x,
\label{eq:H_a1} \\
{\bf H}_a(s)={\bf U}(s)\left(\int_{0}^s {\bf U}^\dagger(x) {\bf W} {\bf U}\mathrm{d}x\right){\bf U}^\dagger(s),
\label{eq:H_a2}
\end{eqnarray}
where $\bf{U}$ is the transformation matrix generated by ${\boldsymbol \eta}$
\begin{equation}
\frac{\mathrm{d} \bf{U}}{\mathrm{d}s}={\boldsymbol \eta}{\bf U}\to{\bf U}(s)=\mathbfcal {T}e^{\int_{0}^s {\boldsymbol \eta(x)\mathrm{d}x}}.
\end{equation}
Equations~(\ref{eq:H_a1}) and (\ref{eq:H_a2}) allow us to estimate ${\bf H}_{a}(s\to\infty)$ and then use
matrix perturbation theory to find the correction to the energy of the eigenstate. We will illustrate this procedure below using the Lieb-Liniger model.

\section{Lieb-Liniger Model}
\label{sec:ll_model}

To test our method, we use the exactly solvable Lieb-Liniger model \cite{lieb1963}, which describes $N$
spinless bosons  on a ring of length $L$. The particles interact via delta functions, so the corresponding one-dimensional Schr{\"o}dinger equation is
\begin{eqnarray}
-\frac{1}{2}\sum_{\alpha=1}^N \frac{\partial^2}{\partial x_{\alpha}^2}\Psi + g\sum_{\alpha_1<\alpha_2} \delta(x_{\alpha_1}-x_{\alpha_2})\Psi=E_N\Psi,
\end{eqnarray} 
where we put $\hbar=m=1$ for convenience. The parameters of the model are $\gamma=g/\rho$ and $e=2E_N/(N\rho^2)$, where $\rho=N/L$
is the density of the system. Since this model is exactly solvable for any $N,L$ and $g$, it gives us a good reference point for testing our approach. 
Note, however, that we do not expect our approach to work extremely well for large systems, as strong correlations preclude the existence of a ``true''
BEC in one spatial dimension. 

\begin{figure}
\centering
\includegraphics[scale=0.9]{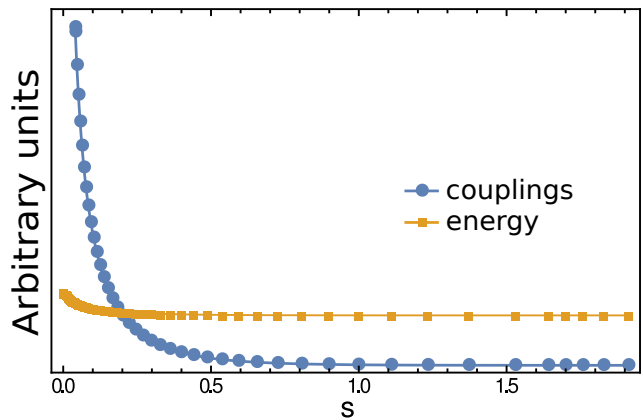}
\caption{The schematic representation of the flow generated by the operator ${\boldsymbol \eta}$ from~(\ref{eq:generator_LL}).
  The circles represent the sum of coupling terms to the $\langle \Phi|{\bf H}| \Phi\rangle$ element, which vanish during the flow.
  The squares show the evolution of $\langle \Phi|{\bf H}| \Phi\rangle$, which is decoupled from the rest at $s\to\infty$. }
\label{Fig:coupl}
\end{figure}

To write the initial matrix elements and the reference state, we use the one-body basis of plane waves, i.e.,
$\phi_i(x)=e^{i k_i x}/\sqrt{L}$, where $k_i \in \{0,\pm 1,\pm 2,...\}2\pi/L$ and $\Phi=L^{-N/2}$. Inspired by the discussion in
section~\ref{sec:prel}, we write the generator as
\begin{equation}
{\boldsymbol \eta}(s)=f_{i0}(s):{\boldsymbol a}_i^\dagger {\boldsymbol a}_0: + \frac{1}{2}\Gamma_{ij00}(s):{\boldsymbol a}_i^\dagger {\boldsymbol a}_j^\dagger {\boldsymbol a}_0 {\boldsymbol a}_0: - H.c..
\label{eq:generator_LL}
\end{equation}
 Here we explicitly relate the parameters of the generator~(\ref{eq:ham_eta}) to the parameters of the Hamiltonian (\ref{eq:eta_ham}) for every $s$.
This generator decouples the element $\langle \Phi|{\bf H}| \Phi\rangle$ from the rest; see figure~\ref{Fig:coupl}, where we plot $\epsilon(s)$ and
$\sum_{p>0} |\langle \Phi_p |{\bf H}(s)|\Phi \rangle|^2$ with $\Phi_p$ containing one- and two-body excitations. Therefore,
the latter represents the coupling to the state of interest. We see that during the flow the couplings vanish, and
$\epsilon(s\to\infty)$ can be interpreted as the eigenvalue of the matrix.  Note that we do not plot any numbers on the $y$
axis as this schematic plot is representative for all considered cases.
 In our code we use units with $L=2\pi$,
  which gives a particularly simple form of the momenta, and sets the scale for the energy and $s$ in the problem. For example,
  the energy difference between the two lowest non-interacting states is one, and therefore the slowest dynamics in the weakly interacting case are described approximately by $e^{-s}$.
  Figure~\ref{Fig:coupl} shows that in these
  units the decoupling indeed occurs for $s$ of ${\cal O}(1)$ as expected.
The study of the flow for other generators is beyond the scope of the present paper.  However, we did check (see~\ref{app:wh_gen}) that the results
obtained with the generator (\ref{eq:generator_LL}) agree with the results obtained using White's generator \cite{hergert2016, white2002},
which includes additional energy denominators compared to~(\ref{eq:generator_LL}).

\subsection{Results}

\begin{figure}
\centering
\includegraphics[scale=0.9]{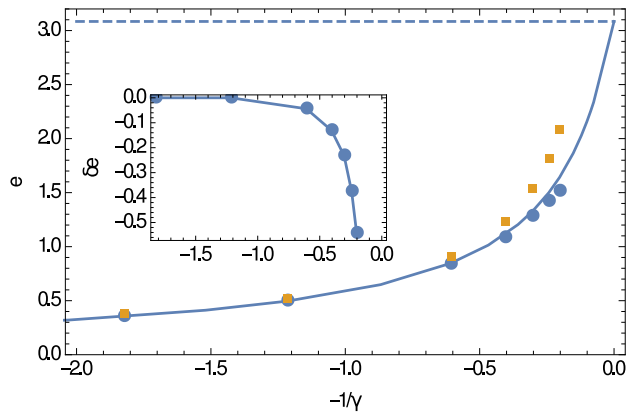}
\caption{The ground state energy per particle, $e=2E_N/(N\rho^2)$,  of the Lieb-Liniger model for $N=4$ as a function of the inverse interaction strength $-1/\gamma$. The solid blue line shows the exact result~\cite{sakmann2005}. The yellow squares are the outcomes of the SRG. The blue circles additionally include the correction $\delta e$. The dashed line represents the ground state energy in the strong coupling limit, i.e., $e(1/\gamma=0)$. The inset shows the behavior of the correction as a function of $-1/\gamma$, the solid line is plotted here to guide the eyes.}
\label{Fig:4part}
\end{figure}

$N=4$. We start with $N=4$. Note that for a cutoff $n\simeq 25$ (in the one-body sector this corresponds to the maximal energy of $288\pi^2/L^2$),
we can easily run the flow until the states are decoupled with high accuracy. Therefore, we have only two sources of error.  The first is due to
the truncation of the Hamiltonian at $s=0$. This error vanishes in the limit of large $n$, but  since the delta function potential
has a hard core (it couples all plane waves equally strongly) the convergence to the $n\to \infty$ limit might be relatively slow (see~\ref{app:conv_two_body}). However, one can still extract accurate results either by fitting 
(see~\ref{app:conv_two_body}) or by using an effective interaction (see~\ref{app:eff_int}).  To be on the safe side, we first solve the problem using the former method and then using the latter. The results of both methods agree well. This is demonstrated explicitly in figures~\ref{fig:convN4gamma3} and~\ref{fig:convN15gamma2} for two parameter sets.

The second error is due to the truncation of the three-body term in Eq.~(\ref{eq:three_body_term}). To estimate this error,
we note that according to~(\ref{eq:H_a1}) for a weak interaction ${\bf H}_{a}\simeq\int \bf{W} \mathrm{d}s$. By definition, the operator $\bf{W}$ connects the state of interest to the states with three excitations. To calculate the contribution to the energy of the perturbation ${\bf H_a}$, we use the standard second-order eigenvalue correction from perturbation theory, i.e.,
\begin{equation}
\delta e \simeq \frac{1}{N}\sum_{p}\frac{\left(\langle \Phi_p| \int_0^\infty {\bf W}(s)\mathrm{d}s | \Phi \rangle\right)^2}{ \langle \Phi | {\bf H} | \Phi \rangle-\langle \Phi_p | {\bf H} | \Phi_p \rangle },
\label{eq:estimators}
\end{equation}
where the sum goes over the all states that contain three particles excited out of the condensate. For consistency, we will keep only the lowest
terms in $g$ in the denominator. 

We show our results in figure~\ref{Fig:4part}. On the scale
  of the figure, the results for the bare delta-function interaction and the effective interaction are indistinguishable.
We see that the SRG reproduces the exact results at weak and moderate coupling strengths. However, when
the interaction strength increases the energy starts to deviate noticeably. This behavior can be understood by calculating $\delta e$. We see that
this term grows very rapidly (numerical analysis reveals that in the considered interval this term grows faster than $\gamma^{2}$) and already at
$\gamma = \pi^2/2$ it accounts for about $25$\% of the SRG result. This shows that the used truncation scheme is not accurate for this $\gamma$ making us stop our calculations. 

\begin{figure}
\centering
\includegraphics[scale=0.9]{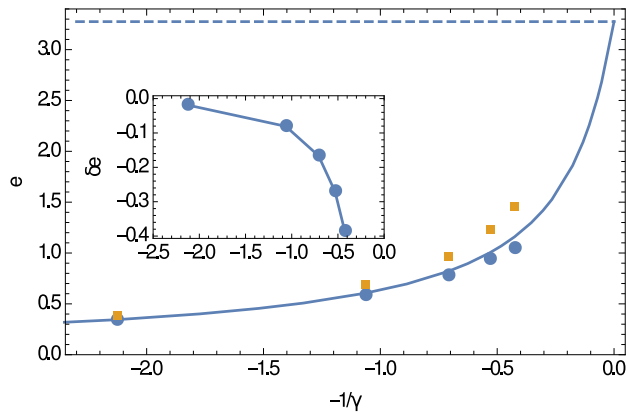}
\caption{The ground state energy per particle,  $e=2E_N/(N\rho^2)$, of the Lieb-Liniger model for $N=15$ as a function of the inverse interaction strength $-1/\gamma$. The solid blue line shows the exact result~\cite{sakmann2005}. The yellow squares are the outcomes of the SRG. The blue circles additionally include the correction $\delta e$. The dashed line represents the ground state energy in the strong coupling limit, i.e., $e(1/\gamma=0)$. The inset shows the behavior of the correction as a function of $-1/\gamma$, the solid line is plotted here to guide the eyes.}
\label{Fig:15part}
\end{figure}

$N=15$.  Our results for $N=15$ are shown in
  figure~\ref{Fig:15part}. On the scale of the figure the results  for the bare delta-function interaction and the effective interaction are again indistinguishable. We see
a similar trend as for $N=4$: The SRG reproduces well the exact results at small and moderate coupling strength,
but fails to describe strongly interacting systems.  The window of applicability of the SRG for $N=15$ is slightly smaller than for $N=4$, which is expected from our error estimation which shows that $\delta e$ grows with $N$ (see~\ref{app:window}).

\section{Conclusions}
\label{sec:concl}

In this paper we have developed a non-perturbative numerical procedure to address bosonic systems with a macroscopic occupation of one state. The method is based on the SRG approach in which 
the Hamiltonian is transformed to decouple the state of interest from the rest.
This transformation is done through a sequence of infinitesimally small rotations in the state space described by a system of differential equations. To make this system solvable with the standard numerical software, we truncate it at the level of three-body operators, and present means to estimate the introduced uncertainty. To illustrate our approach we 
turn to the Lieb-Liniger model, which shows that our flow equations describe small systems with weak and moderate interactions well. Note that our method can be used to describe two- and three-dimensional systems and we use here a one-dimensional model  because its exact solutions allow us to directly test our procedure (although studies of trapped systems in one spatial dimension are interesting on their own right, see~\cite{zinner2016} and references therein) . 

Our approach will allow one to study properties of trapped bosons, systems with a static or mobile impurities~\cite{volosniev2017}. Also, it will be interesting to investigate three-dimensional bosonic bound clusters that appear in different branches of physics such as $^4\mathrm{He}$-clusters in condensed matter physics~\cite{toennies2004} and $\alpha$-clusters in nuclear physics~\cite{Tohsaki2001, freer2012}. In these cases one might need to pick the basis carefully to reduce numerical effort. For instance, if the system is spherically symmetric then the basis should be chosen accordingly (cf.~Ref.~\cite{hergert2016}). 

With some modifications our method can be used to study other set-ups.   In particular, we believe that it is possible to extend the method to bosonic systems without a condensate. To this end, one shall simply follow the steps presented above. First a reference state is used to normal order the operators. This reference state should describe an eigenstate of the Hamiltonian ``well'', such that higher-body excitations are suppressed. As in the present work, the normal ordering provides one with means to truncate the differential equations, opening up the opportunity to approach $N$-body problems using a few-body machinery. Note that a suitable reference state in one-dimensional systems can be obtained by a linear superposition of weakly- and strongly-interacting states~\cite{molte2016}, providing one with a good starting point for this investigation.

\ack
We thank Achim Schwenk and Nikolaj Zinner for useful conversations. We thank Oleksandr Marchukov and Marcel Schmidt for their comments on the video abstract.
A.~G.~V. gratefully acknowledges the support of the Humboldt Foundation. H.-W. H. was supported in part by the
Deutsche Forschungsgemeinschaft through SFB 1245 and
by the German Federal Ministry of Education and Research under contract 05P15RDFN1.

\appendix

\section{Evaluation of commutators}
\label{app:comm}

To write down the flow equations, we need the commutators of the terms in ${\boldsymbol\eta}$ and ${\boldsymbol H}$. For the commutator of one-body operators and one- and two-body operators,
we find:
\begin{eqnarray}
\fl \lambda_{ik} \beta_{mn}  [:{\boldsymbol a}_i^\dagger {\boldsymbol a}_k:, :{\boldsymbol a}_m^\dagger {\boldsymbol a}_n:] = \; (\lambda_{il}\beta_{ln}-\beta_{il}\lambda_{ln})(:{\boldsymbol a}_i^\dagger {\boldsymbol a}_n: + \rho_{in} \mathbf{I}), 
\\ 
\fl\lambda_{ik} \beta_{mnpf} [:{\boldsymbol a}_i^\dagger {\boldsymbol a}_k:, :{\boldsymbol a}_m^\dagger {\boldsymbol a}_n^\dagger {\boldsymbol a}_p {\boldsymbol a}_f:] =  
\; 4\kappa N (\beta_{i00l}\lambda_{lf}-\lambda_{il}\beta_{l00f})(:{\boldsymbol a}_i^\dagger {\boldsymbol a}_f:+\rho_{if}\mathbf{I}) \nonumber \\
\fl \qquad +  2 (\lambda_{il}\beta_{lnpf}-\beta_{inpl}\lambda_{lf})\left[:{\boldsymbol a}_i^\dagger {\boldsymbol a}_n^\dagger {\boldsymbol a}_p {\boldsymbol a}_f:+\rho_{inpf}\mathbf{I}+\kappa P_{inpf}\rho_{np}:{\boldsymbol a}_i^\dagger {\boldsymbol a}_f:\right] \nonumber \\ 
\fl \qquad
= \; 2N\left(2\kappa N -N+1\right)(\beta_{000l}\lambda_{l0}-\lambda_{0l}\beta_{l000})\mathbf{I} + 4\kappa N(\lambda_{0l}\beta_{lif0}-\beta_{0ifl}\lambda_{l0}):{\boldsymbol a}_i^\dagger {\boldsymbol a}_f:\nonumber\\
\fl \qquad + 2 (\lambda_{il}\beta_{lnpf}-\beta_{inpl}\lambda_{lf}):{\boldsymbol a}_i^\dagger {\boldsymbol a}_n^\dagger {\boldsymbol a}_p {\boldsymbol a}_f:,
\end{eqnarray}
here we assume that $\beta_{\alpha_1\alpha_2\alpha_3\alpha_4}=\beta_{\alpha_2\alpha_1\alpha_3\alpha_4}=\beta_{\alpha_1\alpha_2\alpha_4\alpha_3}$,
the same will be assumed for $\lambda_{\alpha_1\alpha_2\alpha_3\alpha_4}$.
Note that with our definition of $\kappa=(N-1)/(2N)$ the element proportional to $\mathbf{I}$ in the second last row vanishes. For the commutator of the two-body
operators, we find:
\begin{eqnarray}
 \fl \lambda_{iklm}\beta_{abcd}[:{\boldsymbol a}_i^\dagger {\boldsymbol a}_k^\dagger {\boldsymbol a}_l {\boldsymbol a}_m:, & :{\boldsymbol a}_a^\dagger {\boldsymbol a}_b^\dagger {\boldsymbol a}_c {\boldsymbol a}_d:] = \nonumber\\
&\; 8\kappa N (\beta_{i00l}\lambda_{lnpf}-\lambda_{inpl}\beta_{l00f} - \lambda_{i00l}\beta_{lnpf}+\beta_{inpl}\lambda_{l00f}):{\boldsymbol a}_i^\dagger {\boldsymbol a}_n^\dagger {\boldsymbol a}_p {\boldsymbol a}_f:\nonumber\\
  &+ 16\kappa N^2\left(\kappa N -N+1\right)(\lambda_{000l}\beta_{l000}-\beta_{000l}\lambda_{l000})\mathbf{I}\nonumber\\
  &+2 N(N-1)(\lambda_{00fl}\beta_{fl00}-\beta_{00fl}\lambda_{fl00})\mathbf{I}  \nonumber \\ 
  &+16\kappa^2 N^2(-\lambda_{000l}\beta_{lif0}+\beta_{0ifl}\lambda_{l000}+\beta_{000l}\lambda_{lif0}-\lambda_{0ifl}\beta_{l000} \nonumber\\
  &\qquad+ \beta_{i00l}\lambda_{l00f}-\lambda_{i00l}\beta_{l00f}):{\boldsymbol a}_i^\dagger {\boldsymbol a}_f: \nonumber \\ 
  &+2(\lambda_{ikfl}\beta_{flcd}-\beta_{ikfl}\lambda_{flcd}) \left[:{\boldsymbol a}_i^\dagger {\boldsymbol a}_k^\dagger {\boldsymbol a}_c {\boldsymbol a}_d:+4\kappa\rho_{kc}:{\boldsymbol a}_i^\dagger {\boldsymbol a}_d:\right] \nonumber \\
  &+ 4(\lambda_{iklj}\beta_{jbcd}-\beta_{iklj}\lambda_{jbcd}){\boldsymbol a}_i^\dagger {\boldsymbol a}_k^\dagger {\boldsymbol a}_b^\dagger {\boldsymbol a}_l {\boldsymbol a}_c {\boldsymbol a}_d.
\end{eqnarray}
The last term proportional to ${\boldsymbol a}_i^\dagger {\boldsymbol a}_k^\dagger {\boldsymbol a}_b^\dagger {\boldsymbol a}_l {\boldsymbol a}_c {\boldsymbol a}_d$ does not
fit in our approximation scheme and should be truncated. Our implementation of this truncation is discussed in~\ref{app:trunc}.

\section{Truncation of the three body operator}
\label{app:trunc}
To truncate the three-body operator, we assume that the number of particles in the lowest state is large, and thus the main
contribution to the ground state energy is due to the piece of ${\boldsymbol a}_i^\dagger {\boldsymbol a}_k^\dagger
{\boldsymbol a}_b^\dagger {\boldsymbol a}_l {\boldsymbol a}_c {\boldsymbol a}_d$ which contains at least one operator $a_0^\dagger$ and
one operator $a_0$. Because of the presence of the condensate, these operators are then treated as numbers, i.e.,
\begin{eqnarray}
4(\lambda_{iklm}\beta_{mbcd}-\beta_{iklm}\lambda_{mbcd}){\boldsymbol a}_i^\dagger {\boldsymbol a}_k^\dagger {\boldsymbol a}_b^\dagger {\boldsymbol a}_l {\boldsymbol a}_c {\boldsymbol a}_d 
\simeq 4 (N-2){\boldsymbol a}_i^\dagger {\boldsymbol a}_j^\dagger {\boldsymbol a}_k {\boldsymbol a}_l L_{ijkl},
\end{eqnarray}
where
\begin{eqnarray}
 \fl  L_{\alpha_i\alpha_j\alpha_k\alpha_l}= &(\lambda_{\alpha_i\alpha_j0m}\beta_{m0\alpha_k\alpha_l}-\beta_{\alpha_i\alpha_j0m}\lambda_{m0\alpha_k\alpha_l})(1+\delta_{\alpha_i0}\delta_{\alpha_j0}-2\delta_{\alpha_j0})(1+\delta_{\alpha_k0}\delta_{\alpha_l0}-2\delta_{\alpha_k0}) \nonumber \\
  &+ (\lambda_{\alpha_i\alpha_j\alpha_km}\beta_{m00\alpha_l}-\beta_{\alpha_i\alpha_j\alpha_km}\lambda_{m00\alpha_l})(1+\delta_{\alpha_i0}\delta_{\alpha_j0}-2\delta_{\alpha_j0})(2-\delta_{\alpha_l0}) \nonumber \\
  &+(\lambda_{\alpha_i00m}\beta_{m\alpha_j\alpha_k\alpha_l}-\beta_{\alpha_i00m}\lambda_{m\alpha_j\alpha_k\alpha_l})(1+\delta_{\alpha_k0}\delta_{\alpha_l0}-2\delta_{\alpha_k0})(2-\delta_{\alpha_i0}) \nonumber \\
  &+(\lambda_{\alpha_i0\alpha_k m}\beta_{m\alpha_j0\alpha_l}-\beta_{\alpha_i0\alpha_km}\lambda_{m\alpha_j0\alpha_l})(2-\delta_{\alpha_i0})(2-\delta_{\alpha_l0}).
\end{eqnarray}

\section{Convergence of the two body energy}
\label{app:conv_two_body}

\begin{figure}
\centering
\includegraphics[scale=0.9]{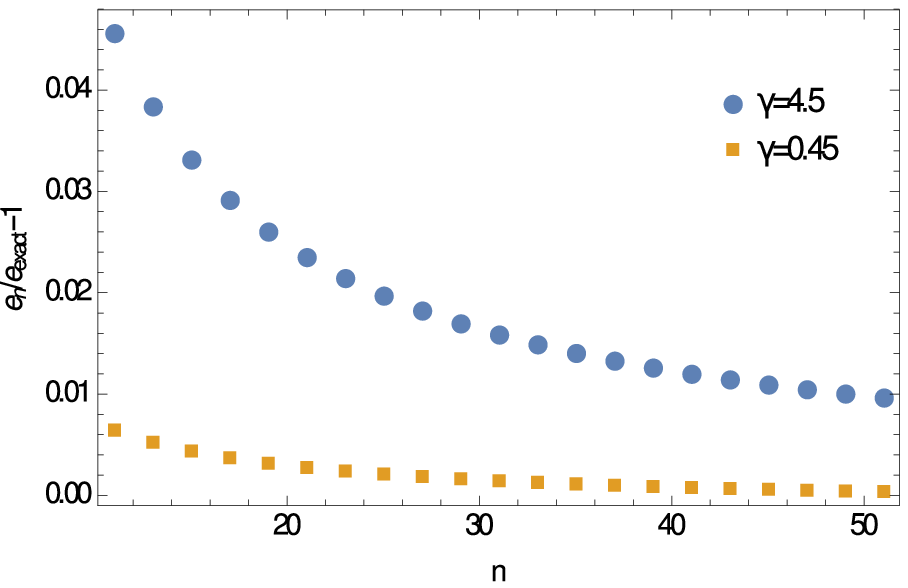}
\caption{The relative error in the energy $(e_n-e_{exact})/e_{exact}$ as a function of the one-body truncation $n$ for the two-body systems with $\gamma=0.45$ (bottom) and $\gamma=4.5$ (top).}
\label{fig:conv2}
\end{figure}

The delta function interaction leads to a cusp in the wave function at zero separation of particles. This non-analyticity implies that accurate results for observables can be
obtained only with a large number of plane wave states. We illustrate this statement by plotting the convergence of the ground state energy versus
the number of the one-body basis states for the Lieb-Liniger model with just two particles, see figure~\ref{fig:conv2}. This plot shows that even in the two-body system
the convergence with $n$ is very slow if $g$ is large. For large $n$ the convergence pattern in the figure can be well approximated by
\begin{equation}
E\simeq E_\infty+\frac{A}{n},
\label{eq:convergence_to_inf}
\end{equation}
where $A$ is some constant that depends on $g$. Note that this convergence is faster than in a harmonic oscillator~\cite{tolle2013, grining2015}
where it is described by $\sim 1/\sqrt{n}$. As is apparent from the discussion below this difference is connected
to a slower growth of the energy with $n$ in a harmonic trap compared to a ring.

To understand this convergence pattern let us 
assume that we have diagonalized the matrix for some cutoff $n$, and obtained the energy $E_n$
and the wave function $\Psi_n$. Now let us see what happens when we diagonalize the Hamiltonian for $n+2$. The corresponding matrix includes the matrix for $n$ coupled to the rest via 
\begin{equation}
\frac{g}{L}\int e^{\frac{2i\pi(n_1x_1+n_2x_2)}{L}}\delta(x_1-x_2)\Psi_n(x_1,x_2)\mathrm{d}x_1\mathrm{d}x_2,
\end{equation}
where at least one of the states $n_1$, $n_2$ was not included in the matrix for $n$.
We have assumed that $n$ is large so $\Psi_n\simeq\Psi_{\infty}$. The function $\Psi_\infty(x_1,x_1)\equiv \bar\Psi$ is constant due to the rotational symmetry of the ring, and therefore we have  
\begin{equation}
\frac{g}{L}\int e^{\frac{2i\pi(n_1x_1+n_2x_2)}{L}}\delta(x_1-x_2)\Psi_n(x_1,x_2)\mathrm{d}x_1\mathrm{d}x_2\simeq g\bar\Psi\delta_{n_1+n_2,0}.
\end{equation}
Now using the second order correction from matrix perturbation theory we calculate the correction to $E_n$ due to the increase of the matrix size 
\begin{equation}
E_{n+2}\simeq E_{n}-\frac{4|g L\bar\Psi|^2}{\pi^2 n^2}.
\end{equation}
Summing the contributions for different $n$ up to infinity, 
this equation leads directly to~(\ref{eq:convergence_to_inf}). In general the leading order correction proportional to $n^{-\delta}$ is characteristic for delta function interactions and we can use it to obtain accurate results from the convergence pattern.  We have observed that for larger number of particles the convergence behavior is also well described by~(\ref{eq:convergence_to_inf}).

\section{Effective Interaction}
\label{app:eff_int}

Another way to produce accurate results for the Lieb-Liniger model is to use some effective potential that reproduces low-energy properties of the system.
For relevant studies of cold atomic systems see references~\cite{rotureau2013, lindgren2014}.  
To introduce this potential, we first notice that all that we need to know about the interaction in our formalism is the following matrix element 
\begin{eqnarray}
\fl \frac{1}{L^2}\int e^{\frac{2\pi i (n_1 x_1 +  n_2 x_2)}{L}} V(x_1-x_2) e^{-\frac{2\pi i (n_3 x_1+ n_4 x_2)}{L}} \mathrm{d}x_1 \mathrm{d}x_2= \nonumber \\ \frac{\delta_{n_1+n_2,n_3+n_4}}{L}\int e^{2\pi i x(n_1-n_3)} V(x) \mathrm{d}x \equiv \frac{\delta_{n_1+n_2,n_3+n_4}}{L} V_{n_1n_3} ,
\end{eqnarray}
where $|n_i|\leq n_{max}$ and $n_{max}$ is the truncation parameter defined by $n$. 
Apparently such a matrix element also appears when we solve the Schr{\"o}dinger equation
in the 'relative' coordinates
\begin{equation}
-\frac{\partial^2}{\partial x^2}\psi + V(x)\psi=E\psi,
\end{equation}
by expanding the wave function $\psi$ in the plane wave basis, i.e., $\psi=\frac{1}{\sqrt{L}}\sum_{|n_l|\leq n_{max}} a_l e^{-2\pi i n_l x/L}$ and solving the matrix equation
\begin{equation}
(V_{\alpha_1j}+T_{\alpha_1 j}){\mathcal Q}_{j\alpha_2}={\mathcal Q}_{\alpha_1 j}E_{j \alpha_2}.
\end{equation}
Here $\mathbfcal{Q}$ is the matrix that contains eigenvectors as columns, ${\bf E}$ is the diagonal matrix that contains the eigenvalues, and $\mathbf{T}$ is the kinetic energy. Now we can turn the question around and
find the potential that within our truncation space gives some specific matrices $\mathbf{E}$ and $\mathbfcal{Q}$. Such the potential then reads
\begin{equation}
\mathbf{V}=\mathbfcal{Q}\mathbf{E}\mathbfcal{Q}^T-\mathbf{T}.
\end{equation} 

Let us now specify the desired low-energy properties.
First of all, we fix the energies $E_{\alpha \alpha}$ to the $n$ lowest eigenenergies of the equation
\begin{equation}
-\frac{\partial^2}{\partial x^2}\psi_\alpha + g\delta(x)\psi_\alpha=E_{\alpha\alpha} \psi_\alpha,
\end{equation}
this choice means that in the two-body sector we always obtain correct energies. 
Next, we define the matrix $\mathbfcal{Q}$ as 
\begin{equation}
\mathbfcal{Q}=\frac{1}{\sqrt{{\bf u}^T {\bf u}}} \bf{u},
\end{equation}
where the matrix ${\bf u}$ is an $n\times n$ matrix defined as $u_{\alpha_1 \alpha_2}=\frac{1}{\sqrt{L}}\int_{0}^{L}e^{ 2\pi i \alpha_1 x/L}\phi_{\alpha_2}(x)$. We see that if $n\to\infty$ then ${\bf u}^T {\bf u} \to 1$, and we have $\mathbfcal{Q}\to {\bf u} $. Therefore, the matrix ${\mathbfcal Q}$  is an orthogonal matrix that approximates the eigenstates and for $n\to\infty$ it reproduces the exact results.

\begin{figure}
\centering
\includegraphics[scale=0.9]{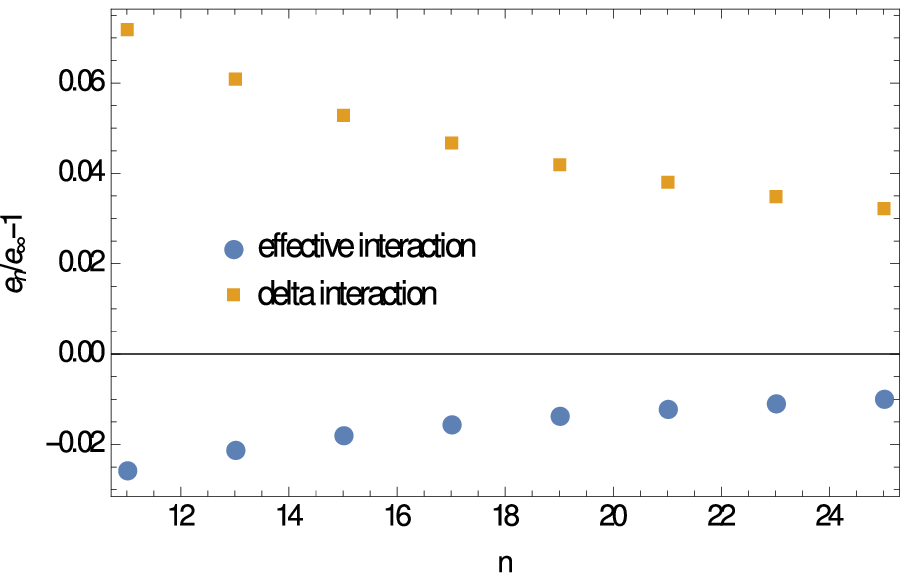}
\caption{The relative error in energy $e_n/e_{\infty}-1$ as a function of the one-body truncation $n$. The parameters are $N=4$, $L=2\pi$, and $\gamma = \pi^2/3\sim 3.3$. The value $e_{\infty}$ is obtained from the fit $e_n=e_{\infty}+c n^{-\delta}$, where $e_{\infty}, c,\delta$ are the fitting parameters. The values $e_\infty$ obtained for the effective interaction and the delta potential differ by less than 0.01 {\%}.} 
\label{fig:convN4gamma3}
\end{figure}

\begin{figure}
\centering
\includegraphics[scale=0.9]{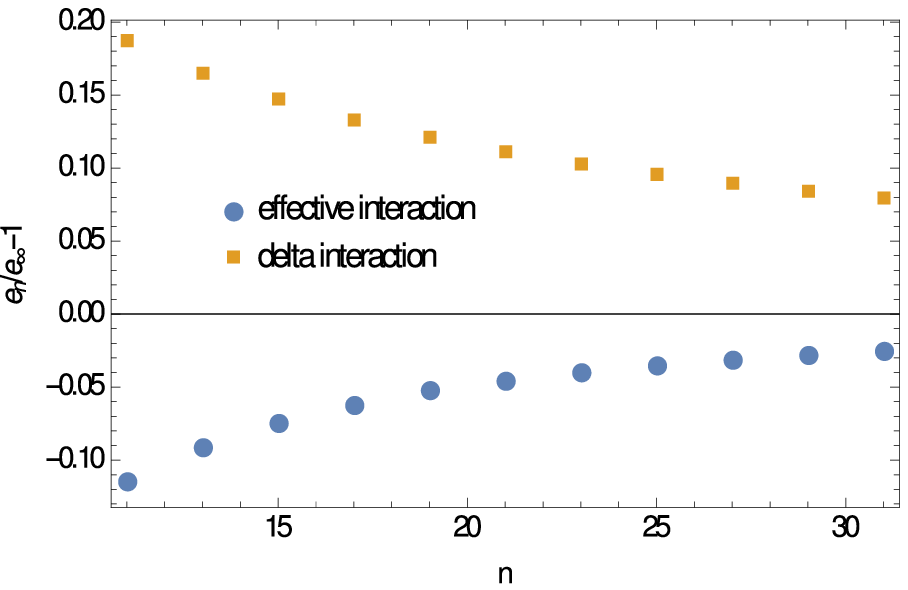}
\caption{The relative error in energy $e_n/e_{\infty}-1$ as a function of the one-body truncation $n$. The parameters are $N=15$, $L=2\pi$, and $\gamma = 5\pi^2/21\sim 2.35$. The value $e_{\infty}$ is obtained from the fit $e_n=e_{\infty}+c n^{-\delta}$, where $e_{\infty}, c,\delta$ are the fitting parameters. The values $e_\infty$ obtained for the effective interaction and the delta potential differ by less than 1 {\%}.}
\label{fig:convN15gamma2}
\end{figure}

The effective interaction shows faster convergence than the zero-range interaction, see figures~\ref{fig:convN4gamma3} and \ref{fig:convN15gamma2}, which depict our results for a few representative cases.  By comparing the fitted values for the zero-range interaction and for the effective interaction we cross-check the two methods and insure accuracy of our results. The convergence pattern for the delta function potential is usually well described by~\ref{eq:convergence_to_inf}. Note that we cannot directly apply the same line of arguments to find the convergence pattern for the effective interaction potential. Indeed, in this case the increase of the matrix size leads to a change of all matrix elements, and, therefore, standard perturbation theory cannot be used.


\section{Other generators}
\label{app:wh_gen}

In section~\ref{sec:prel}, we present examples of different generators that can be used to create the flow, see also~\cite{wegner1994, Tsukiyama2011,hergert2016, white2002, kehrein2006}. In the main text, we illustrate our method using exclusively the operator~(\ref{eq:generator_LL}) and leave other generators for future studies. Note that other ${\boldsymbol \eta}$ can be used directly in the derived flow equations (\ref{eq:flow_eq_bosons1})-(\ref{eq:flow_eq_bosons3}) after the parameters of the generator~(\ref{eq:ham_eta}) are specified.  In this appendix, we briefly discuss the use of the White-type generator
\begin{equation}
{\boldsymbol \eta}^{White}(s)= \xi_{i0}(s):{\boldsymbol a}_i^\dagger {\boldsymbol a}_0: + \frac{1}{2}\eta_{ij00}(s):{\boldsymbol a}_i^\dagger {\boldsymbol a}_j^\dagger {\boldsymbol a}_0 {\boldsymbol a}_0: - H.c.,
\end{equation}
where 
\begin{equation} \xi_{\alpha 0}=\frac{f_{\alpha 0}}{f_{\alpha \alpha}-f_{0 0}}, \qquad \eta_{\alpha_1 \alpha_2 00}=\frac{\Gamma_{\alpha_1 \alpha_2 0 0}}{f_{\alpha_1 \alpha_1}+f_{\alpha_2 \alpha_2}-2 f_{0 0}}.
\end{equation}
 This generator is similar to the one in~(\ref{eq:generator_LL})  but it has additional energy denominators. As can be infered from section~\ref{sec:prel} for weak interactions this leads to the simultaneous decay of all couplings with $e^{-s}$. 

\begin{figure}
\centering
\includegraphics[scale=0.9]{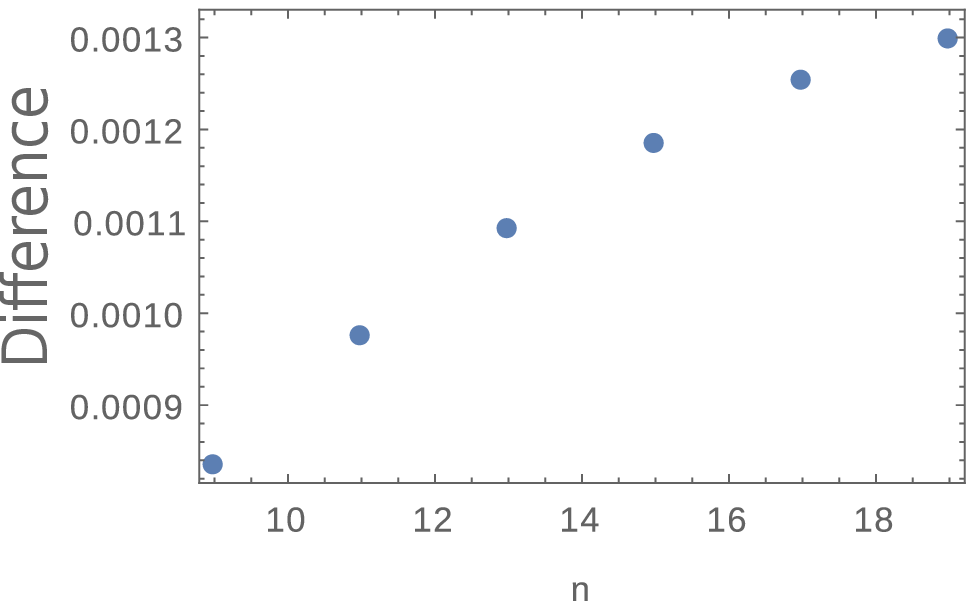}
\caption{ The relative difference $e_n/e^{White}_n-1$ as a function of the one-body truncation $n$. Here $e_n$ is calculated using the generator $\eta$ in~(\ref{eq:generator_LL}),  and $e^{White}_n$ using $\eta^{White}$.  The parameters are $N=4$, $L=2\pi$, and $\gamma = \pi^2/2\sim 4.93$. The fit $e_\infty/e^{White}_\infty-1+c n^{-\delta}$, where $e_\infty/e^{White}_\infty, c,\delta$ are the fitting parameters, leads to $e_\infty/e^{White}_\infty-1\simeq 0.002$.}
\label{fig:white_gen}
\end{figure}

Without truncation, the operators $\eta^{White}$ and $\eta$ in (\ref{eq:generator_LL}) define a unitary transformation and consequently lead to the exact energies. Our truncation scheme spoils this property, but it turns out that for the considered cases the results of the two generators are still very close to each other. We illustrate this statement in figure~\ref{fig:white_gen} for $N=4$ and $\gamma=\pi^2/2$. The correction $\delta e$ for this case accounts for about quarter of the SRG result meaning that the truncation procedure is no longer accurate, still the relative difference between the two results is a fraction of a percent. Therefore, for this problem these two generators can be used interchangeably.  

\section{Dependence of $\delta e$ on $N$.}
\label{app:window}

In our work we noticed that $\delta e$ from~(\ref{eq:estimators}) increases with the number of particles $N$ for a fixed density $n$ and $\gamma$. This feature can be observed in figures~\ref{Fig:4part} and~\ref{Fig:15part} where for same values of $\gamma$ these corrections in the $N=4$ case are smaller than in the $N=15$ case. We report a similar behavior also in~\cite{volosniev2017}.  To understand this growth, let us first analyze the flow equations (\ref{eq:flow_eq_bosons1})-(\ref{eq:flow_eq_bosons3}) with the generator~(\ref{eq:generator_LL}) in the limit $g N \to 0$. In this case $S_{\alpha_1\alpha_2\alpha_3\alpha_4}\ll N S_{\alpha_1\alpha_2\alpha_3\alpha_4\alpha_5 \alpha_6}$, and, thus, 
\begin{eqnarray}
\frac{\mathrm{d}\Gamma_{\alpha_1\alpha_2\alpha_3\alpha_4}(s)}{\mathrm{d}s}\simeq \frac{(1+P_{\alpha_1\alpha_2})(1+P_{\alpha_3\alpha_4})}{2}S_{\alpha_1\alpha_2\alpha_3\alpha_4}.
\end{eqnarray}
We can estimate $\Gamma_{\alpha_1\alpha_2\alpha_3\alpha_4}$ using $f_{\alpha_1 \alpha_2}(s)$ instead of $f_{\alpha_1 \alpha_2}(0)$ in $S_{\alpha_1\alpha_2\alpha_3\alpha_4}$. If we do so, we find that $\Gamma$ is simply proportional to $g$, and, hence, $\delta e/e \sim \gamma^4 N^4$. We see that in the limit $g N\to 0$ the correction grows very rapidly with $N$. 

\begin{figure}
\centering
\includegraphics[scale=0.9]{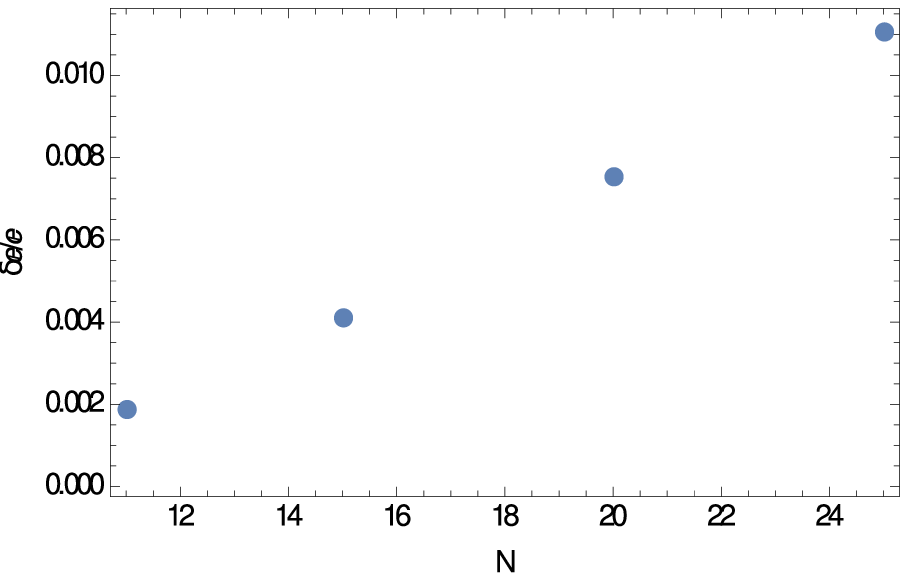}
\caption{ The ratio $\delta e/e$ as a function of $N$ for the Lieb-Liniger model with $\gamma=0.1$. Here $e$ is the energy per particle and the correction $\delta e$ is calculated using~(\ref{eq:estimators}). }
\label{fig:window}
\end{figure}

We are not able to provide a simple analytical analysis if the terms with $N S_{\alpha_1\alpha_2\alpha_3\alpha_4\alpha_5 \alpha_6}$ in~(\ref{eq:flow_eq_bosons3}) are large. Instead, we investigate this case numerically. To this end, we choose to work with $\gamma=0.1$. We find (see figure~\ref{fig:window}) that the ratio $\delta e/e$ increases with $N$, however, slower than $N^4$. Fitting suggests a much milder $\sim N^2$ scaling in this window of $N$.

 \end{document}